# Extreme value statistics and the Pareto distribution in silicon photonics

**D Borlaug, S Fathpour and B Jalali**
Department of Electrical Engineering, University of California, Los Angeles, 90095

Email: jalali@ucla.edu

**Abstract.** L-shape probability distributions are extremely non-Gaussian distributions that have been surprisingly successful in describing the frequency of occurrence of extreme events, ranging from stock market crashes and natural disasters, the structure of biological systems, fractals, and optical rogue waves. In this paper, we show that fluctuations in stimulated Raman scattering in silicon, as well as in coherent anti-Stokes Raman scattering, can follow extreme value statistics and provide mathematical insight into the origin of this behavior. As an example of the experimental observations, we find that 16% of the Stokes pulses account for 84% of the pump energy transfer, an uncanny resemblance to the empirical Pareto principle or the 80/20 rule that describes important observation in socioeconomics.

**1. Introduction**
Extreme value theory is a branch of statistics that deals with the extreme deviations from the median of probability distributions. One class of extreme value statistics are L-shape probability density functions, a set of power-law distributions in which events much larger than the mean (outliers) can occur with significant probability. In stark contrast, the ubiquitous Gaussian distribution heavily favors events close to the mean, and all but forbids highly unusual and cataclysmic events. Extreme value statistics have recently been observed during generation of optical Rogue waves, a soliton-based effect that occurs in optical fibers [1]. Here, we experimentally show that the distribution of Raman amplified pulses in silicon, as well as coherent anti-Stokes scattering (CARS), in the presence of a noisy pump, follow similar power law statistics. We describe a model that shows how such unusual distributions emerge from the interaction of random fluctuations with a strongly nonlinear function.

There have been a number of studies of noise in stimulated Raman scattering (SRS) in optical fibers. These studies have primarily focused on the role of quantum fluctuations in the amplified spontaneous emission in fibers [5-10]. The effects of dispersion and walk off on the noise characteristics of Stokes pulses in fibers have been studied and a slight skew towards the low energy tail of the distribution was observed in simulations [9]. Shaping of spontaneous emission noise by Raman amplification in fibers has been observed where it was shown that the high energy tail of the distribution is actually suppressed due to initiation of 2nd order Raman scattering [10]. An enhancement of the high energy tail of the distribution for amplified spontaneous emission in optical fibers has been reported [11], however, no connection with extreme value statistics was made. With the exception of the latter, in all these cases, the dominant noise source was believed to be quantum fluctuations in the initial spontaneous emission that triggers the stimulated emission.

Stimulated Raman scattering (SRS) in silicon has been extensively pursued as a means to create optically pumped amplifiers and lasers and hence to bypass the material's indirect band structure [12]. The technology is particularly promising in the mid-infrared where two photon absorption, a competing process that also creates free carrier absorption, is absent. In this paper, we describe the first study of the statistics of Raman amplification in silicon. In experiments performed at mid-IR, we show how L-shape distributions arise from the Raman amplification of an input Stokes beam by a noisy pump [13].

## 2. Experiment

The experiments were performed at mid-IR wavelengths on a 2.5 cm thick bulk silicon sample. The pump laser was a Q-switched Nd:YAG pumped OPO emitting pulses at a wavelength of 2.88 microns. The pulse width as reported by the manufacturer is approximately 4 ns (FWHM), but it is known that the envelope has random intra-pulse structure due to multiple longitudinal modes that have pulse-to-pulse fluctuations. The Stokes was a CW HeNe laser at 3.4 microns. After traveling through the sample, the beams are separated using a Dichroic beam splitter. Further isolation between the two beams is achieved using a spectrometer. The pump wavelength was detected with a photodetector (Coherent J25LP-MB) with a response time of 80 microseconds. The Stokes wavelength was detected with an InAs photodetector with a response time of 15 ns. The detectors were not fast enough to resolve the intra-pulse structure of the envelope, however, this is of no concern because we are merely interested in pulse energy fluctuations. Measuring these fluctuations only requires pulse-to-pulse relative energy measurements, for which the detectors are adequate. The detected pump and stokes signals are then measured on an oscilloscope which is triggered at the pump repetition rate. The oscilloscope waveforms are recorded on a remote laptop running labview.

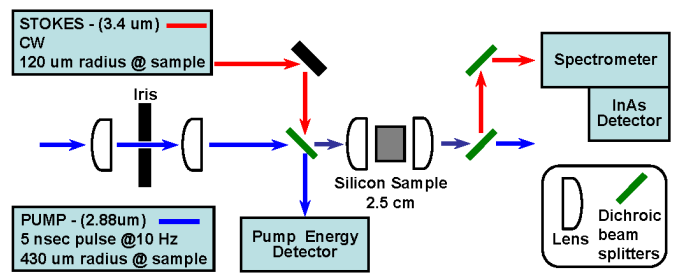

**Figure 1.** Experimental setup.

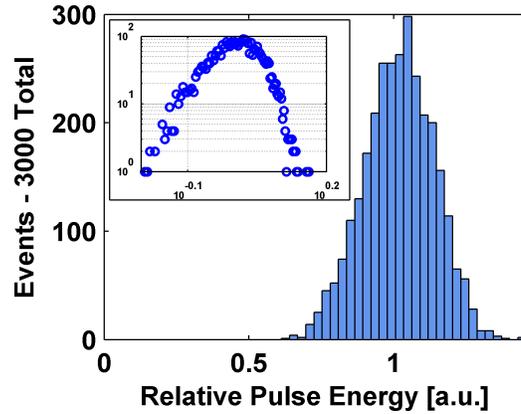

**Figure 2.** Measured histogram of pump pulses. Inset shows the same data plotted on the a log-log scale.

Figure 2 shows the histogram of 3000 pulses for the pump pulse energy. As can be seen, the distribution is nearly symmetric around a well defined mean energy – in other words, it has the general features of an ordinary (Gaussian or Rician) distribution. The Raman gain experienced by the Stokes signal is obtained by comparing the output Stokes peak pulse energy in the presence and absence of the pump. Again, only relative pulse-to-pulse statistical fluctuations are of interest here. The histogram (3000 pulses) of the Raman gain is shown in figure 3.

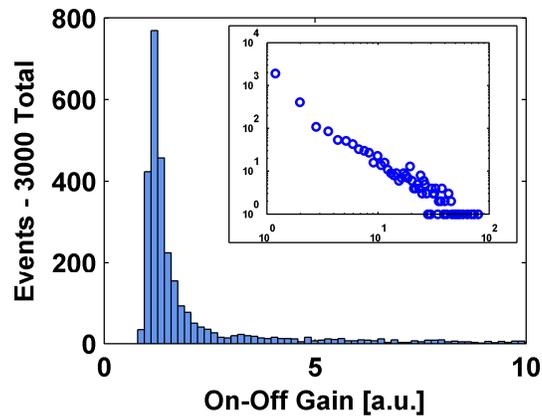

**Figure 3.** Measured histogram of amplified Stokes pulses. Inset shows the same data plotted on a log-log scale. Extreme value statistical behavior and outlier events are clearly evident in the tail of the L-shape distribution.

The Stokes input is the same for all pulses, therefore the gain distribution also describes the distribution of amplified Stokes pulses. The observed distribution clearly shows L-shape extreme value behavior, highlighted by the high probability of large outliers in the "fat-tail" of the distribution. While most pulses experience modest gain, a small percentage of pulses experience amplification far larger than that of average events. This is even more evident in the log scale plot shown in the inset. The distribution shows that the number of such pulses does not decay exponentially with their amplitude, as would be expected for a Gaussian distribution, but rather has a power-law dependence, as expected for an extreme value random process.

It has previously been shown that the Rician distribution [2] is the proper model for amplitude fluctuations arising from a coherent field perturbed by narrowband noise [3-4]. We invoke it here as an approximation to fluctuations of the pump laser. Mathematically, the Rician distribution is given by,

$$f(E_p) = \frac{E_p}{\sigma^2} Exp\left[-\frac{E_p^2 + \upsilon^2}{2\sigma^2}\right] I_0\left[\frac{E_p \upsilon}{\sigma^2}\right] \qquad (1)$$

where $E_p$ is the electric field amplitude for the pump laser, $I_0$ is the modified Bessel function of the first kind with order zero, and $\upsilon$ and $\sigma$ are parameters that influence the offset and width of the distribution. The function approaches a Rayleigh distribution in the limit when $\upsilon \to 0$, evident in figure 4.

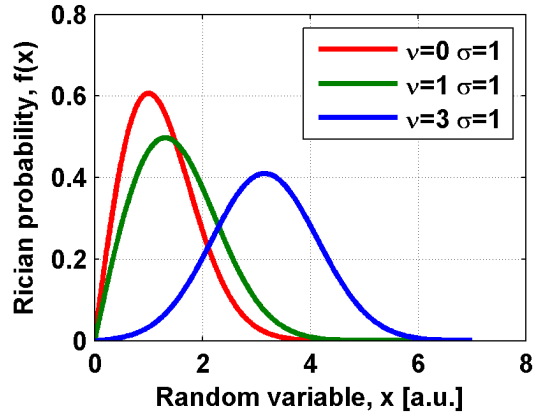

**Figure 4.** The Rician distribution.

Through a nonlinear transformation, the corresponding pump intensity distribution is given by,

$$f(I_p) = \frac{Z_0}{4n\sigma^2} Exp\left[-\frac{Z_0 I_p/2n + \upsilon^2}{2\sigma^2}\right] I_0\left[\frac{\upsilon}{\sigma^2}\sqrt{\frac{Z_0 I_p}{2n}}\right] \qquad (2)$$

Here, $I_p = 2nE_p^2/Z_0$, where $Z_0$ is the impedance of free space and $n$ is the refractive index of the medium.

To show how extreme value behavior emerges from the seemingly innocuous Rician distribution we start with the simple expression for the Raman gain,

$$G = e^{g_R \cdot I_p \cdot L} \qquad (3)$$

Here $g_R$ is the Raman gain coefficient and $L$ is the pump-Stokes interaction length. This nonlinear function causes the pump intensity fluctuations, given by equation 2, to be transformed into the following distribution for the Raman gain,

$$f(G) = G^{-(1+b/2\sigma^2)} \cdot \left(\frac{b}{2\sigma^2} Exp\left[-\frac{\upsilon^2}{2\sigma^2}\right]\right) \cdot I_0\left[\frac{\upsilon\sqrt{bLn(G)}}{\sigma^2}\right] \qquad (4)$$

where $b \equiv Z_0/2ng_R L$. The first term in this expression readily predicts that the Raman gain will have a power law behavior, the hallmark of extreme value statistics. Equation 4 is graphically plotted in figure 5.

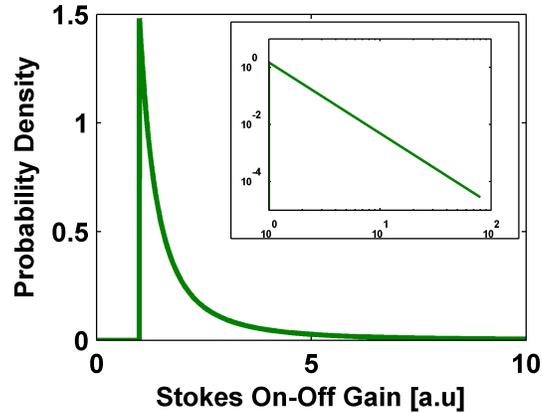

**Figure 5.** Distribution of amplified Stokes pulses according to the model. Inset shows the same function plotted on a log-log scale. The model produces the salient features of experimental data shown in Fig. 3.

The inability to measure the pulse envelope and its intra-pulse structure limits the measurement to that of energy fluctuations, whereas, the model is based on the intensity fluctuations. This prevents a quantitative comparison between the model and experiment. However, the model unambiguously explains the salient feature of the experimental observations. The model shows how extreme value statistics, for the Stokes, emerge through nonlinear transformation of the pump intensity distribution in the case of Raman amplification. Although a Rician distribution was chosen to represent the pump fluctuations, this choice is not critical. Extreme value behavior for the Stokes emerges from Gaussian pump distributions as well.

The same experimental setup, as show in figure 1, is used to measure the statistics of anti-Stokes pulses produced through CARS. This is a parametric process where Stokes photons are converted to anti-Stokes photons through the Raman nonlinearity. The spectrometer is tuned to the anti-Stokes wavelength and the same data acquisition method is used. The anti-Stokes signal is compared to the CW stokes signal entering the gain medium. No attempt was made to create phase matching, therefore the conversion efficiency is expected to below. The anti-Stokes distribution also has extreme value behavior as evidenced by a clear skew towards high energy pulses. However, the extent of the high energy tail of the distribution is less than that observed for Stokes pulses. This can be attributed to the fact the low conversion efficiency effectively reduces the extent of the extreme value tail along the x-axis. The distribution of CARS converted anti-Stokes pulses should more closely mirror those of the Stokes pulses when phase matching condition is satisfied.

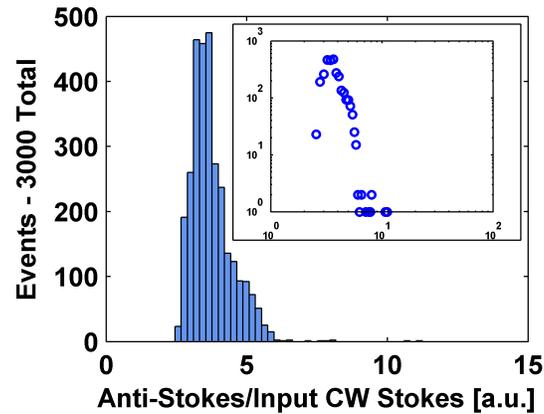

**Figure 6.** Measured histogram of anti-Stokes pulses.

The extreme value behavior reported here has an intriguing association with power law behavior observed in social phenomena. Although the number of extreme value pulses is small, their contribution to the total energy cannot be ignored. Pump pulses with slightly larger power than the mean create un-proportionally larger gain through the nonlinear transfer function of the gain process. In what can be viewed as the "survival of the fittest (pulses)", the impact of high energy pulses in the output Stokes distribution is magnified. In fact, we find that 84% of the pump energy transfer occurs for 16% of the pulses, an intriguing resemblance to the Pareto Principle otherwise known as the "80/20 rule" which contends that 20% of the population controls 80% of the fiscal wealth [14]. Similar behavior is also observed in many other phenomena such as geographical distribution of population and casualty losses from natural disasters. This unlikely parallel between nonlinear optics and these seemingly unconnected systems arises because in both case, the statistics are governed by the transformation of a normal distribution, representing the initial condition, by a nonlinear transfer function.

## 3. Summary

In summary, we have shown that the fluctuations of Raman amplified Stokes signal and coherent anti-Stokes Raman scattered signal, in the presence of a noisy pump, follow extreme value statistics and have provided mathematical insight into its origin. Our observations have important practical implications. For example, a few Stokes pulses carry most of the converted beam energy. Also, the extreme deviation from Gaussian statistics implies that the traditional characterization based on the standard deviation will provide a highly incomplete description of the pulse to pulse fluctuations.

**Acknowledgments**
This work was supported by Dr. Henryk Temkin of DARPA-MTO. The authors thank Drs. D. Dimitropoulos and D.R. Solli of UCLA for discussions.